# Wavelength-agnostic 3D-Nanoprinted coupler

HUIYU HUANG,[1] ZHITIAN SHI,[1] CHUNHUI YAO,[1, 2] RICHARD PENTY,[1] QIXIANG CHENG[1, 2] *

[1]*Electrical Engineering division, Department of Engineering, University of Cambridge, Cambridge, United Kingdom*
[2]*GlitterinTech, Limited, Cambridge, UK*
*\*qc223@cam.ac.uk*

**Abstract:** We present a photonic coupler that exhibits effectively wavelength-agnostic performance for ultra-broadband optical interfacing. By incorporating a dual-ellipsoidal geometry, the design facilitates quasi-free-space optical propagation. We further propose a hybrid modelling workflow employs a matrix optics–based approach as an efficient pre-design tool, capturing critical geometry-to-mode mapping characteristics, significantly narrowing the parameter space required for subsequent full-vectorial finite-difference time-domain (FDTD) simulations. Our design achieves a 1 dB bandwidth exceeding 800 nm coupling from fibre to chip, with an insertion loss as low as 1.3 dB—to the best of our knowledge, a record for any reported photonic couplers. The additive manufacturing approach via 3D nano-printing enables flexible geometry customization and sub-micron integrated alignment features, facilitating seamless integration with photonic chips and optical fibers. Experimental validation demonstrates excellent stability and thermal robustness across diverse operational conditions, highlighting the design's suitability for integration into wide range of broadband photonic systems.

## 1. Introduction

Over the past few decades, photonic integrated circuits (PICs) have emerged as cornerstones of modern photonic technologies [1,2]. While many standardized photonic systems, such as optical transceivers [3] and signal processors [4], typically operate within fixed frequency bands, emerging applications increasingly demand access to significantly broader wavelength ranges. For example, in optical communications, cutting-edge research has begun to explore full-band coverage from the O-band to the U-band, leveraging technologies such as broadband optical frequency combs [6] to achieve unprecedented data throughput. Moreover, in the field of optical sensing, typical near-infrared spectroscopy for biomarker detection or industrial chemical analysis requires several hundreds to over a thousand nanometers of continuous detection bandwidth to resolve distinct molecular features [7, 8]. Serving as the energy transfer interface between photonic chips and optical fibers, photonic couplers play a foundational role in enabling such ultra-broadband operation. However, existing coupler designs remain largely wavelength-dependent, posing a fundamental bottleneck to broadband PIC integration and packaging [9,10]. In-plane coupling generally relies on edge couplers positioned at the chip facets, requiring stringent micron-scale alignment precision and typically exhibiting limited bandwidths to less than 100 nm [11]. Out-of-plane coupling, commonly employing grating couplers, relaxes alignment but further limits achievable bandwidth [12]. In recent years, photonic wire bonding has emerged as a promising technique offering flexible optical coupling interface between different material systems [13]. Nevertheless, all aforementioned coupling techniques depend on confined guided modes or bound optical states, inherently restricting their bandwidth. This limitation underscores an urgent need for a wavelength-agnostic coupler design that is capable of addressing the ever-increasing bandwidth demands of PICs, especially advanced optical sensing, which typically spans over hundreds of nanometers [14].

In this paper, we present a wavelength-agnostic photonic coupler targeting for ultra-broadband optical interfacing, as illustrated in Figure 1. This design employs a dual-ellipsoidal geometry, enabling incident light to undergo two consecutive total internal reflections (TIR) before transitioning to the targeted waveguide. Different to conventional coupling methods that depend on guided modes or adiabatic transitions, our approach harnesses quasi-free-space propagation, inherently eliminating wavelength dependencies while maintaining low insertion loss. Our device achieves an insertion loss as low as 1.3 dB coupling from fiber to chip and maintains a 1 dB bandwidth exceeding 800 nm —to the best of our knowledge, a record for any reported photonic couplers. The coupler is fabricated using two-photon polymerization (TPP), a 3D nano-printing technique providing sub-micron resolution, ideally suited for freeform micro-optics [15]. This additive manufacturing approach additionally allows on-demand customization of coupler geometry, facilitating efficient mode matching between optical fibers and photonic chips without extensive redesigns or lithographic masks, To guarantee alignment tolerance and mechanical robustness, a 3D-coprinted funnel is incorporated enclosing the coupling structure. Extensive experimental evaluation under thermal cycling and high optical intensities have been further demonstrated, verifying its stable and robust coupling characteristics and confirming its suitability for wide-range of applications.

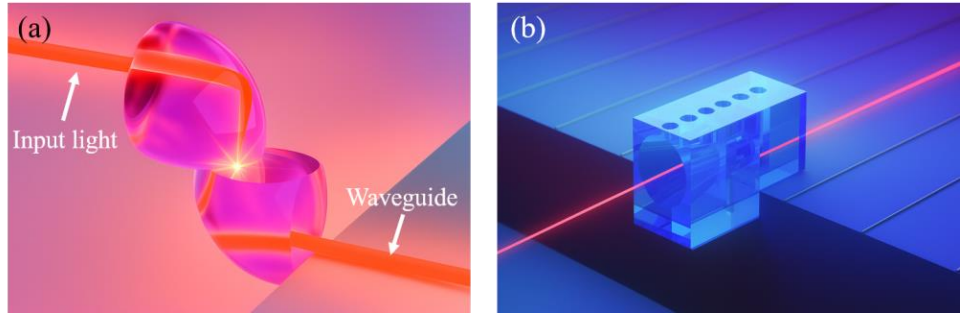

Fig. 1 Schematic views of the devices: (a) On-chip 3D-printed fiber-to-chip coupler; (b) on-chip printed funnel to secure stripped optical fibres.

## 2. Theoretical modelling and FDTD simulation

To achieve ultra-broadband and low-loss coupling, our design leverages a quasi-free-space light propagation facilitated by a pair of 3D-nanoprinted freeform reflectors. After being extracted into the coupler, the light losses its form as a confined mode due to the expanded geometry. This enables quasi-free-space propagation, rendering the coupler intrinsically wavelength-agnostic. However, the inherent asymmetry and extended footprint of the coupler significantly increase computational complexity in full-wave simulations. Conventional optimization methods relying on exhaustive parameter sweeps become impractical, particularly when considering ultra-broad wavelength ranges or diverse fibre modes [16]. To mitigate this computational challenge, we employ a matrix optics–based modelling approach, offering a first-order yet physically intuitive approximation of the optical path within the coupler. By discretizing the optical system into segments governed by ray transfer matrices, representative rays—originating from the waveguide facet—are systematically traced as they reflect and focus toward the fibre mode. This ray-based formulation efficiently captures critical geometry-to-mode mapping characteristics, substantially reducing the necessity for densely populated 3D simulation grids.

### 2.1. Theoretical modelling by matrix optics

The matrix optics model serves as an efficient pre-design tool by rapidly identifying viable reflector geometries, guiding initial parameter selection, and significantly narrowing the parameter space required for subsequent full-vectorial finite-difference time-domain (FDTD) simulations. This hybrid modelling workflow accelerates the design cycle and enhances physical insight without compromising modelling fidelity. Firstly, the incident light beam is approximated as a point source located at the focal point of the elliptical reflector along the y-axis, with a defined divergence angle, as illustrated in Figure. 2 (a). According to ISO 11146-1:2005, for a Gaussian beam emitted from an optical fibre facet entering the reflecting units, the divergence half-angle $\theta$ is expressed by:

$$\theta = \frac{\lambda}{\pi n w_0} \quad (1)$$

where $\lambda$ is the wavelength, $n$ is the refractive index of the resin, and $w_0$ is the beam waist size. A similar estimation applies to integrated waveguides. For example, a laser beam at 1310 nm from a 980HP optical fibre with an MFD of approximately 5.71 μm yields a divergence half-angle of about 10.47°. Following the initial approximation, the dual-reflection optical process within the coupler can be effectively described using matrix optics. As illustrated in Figure. 2 (a), the approximated point source undergoes two total internal reflections (TIR) at the air-polymer interface, allowing the propagation path within the reflectors to be divided into seven distinct segments: two transmissions at the fibre-reflector interfaces, three free-space propagations, and two TIR events. These optical segments each introduce characteristic transformations to the propagating rays, which can be mathematically described by individual ray-transfer matrices.

If we represent the incident light beam by the vector $\begin{bmatrix} x \\ \theta \end{bmatrix}$, the output beam vector $\begin{bmatrix} x' \\ \theta' \end{bmatrix}$ is given by:

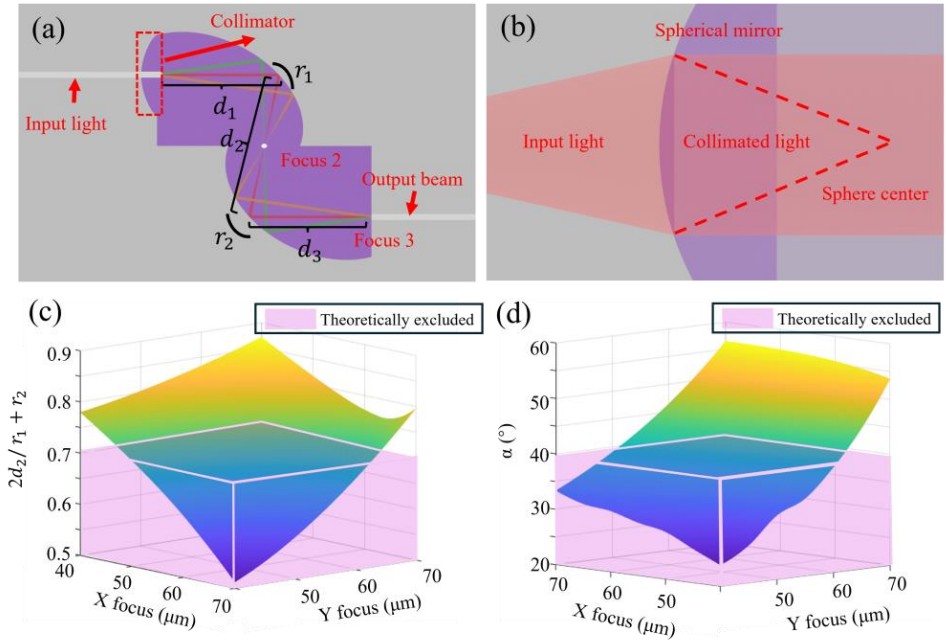

Fig. 2 (a) Illustration of the light propagation within the reflector system. Incident light undergoes total internal reflections (TIR) at the air-polymer interface, subsequently converging into a collimated beam at the output facet; (b) Planar schematic representation of the spherical collimation lens geometry; (c) Theoretical depiction of an aberration-free beam profile, with areas enclosed by the red dashed lines indicating regions theoretically excluded to achieve ideal performance; (d) Geometric constraints imposed by total internal reflection (TIR) conditions, highlighting regions eliminated from design consideration, as indicated by areas enclosed with red dashed lines.

$$\begin{bmatrix} x' \\ \theta' \end{bmatrix} = \begin{bmatrix} A & B \\ C & D \end{bmatrix} \begin{bmatrix} x \\ \theta \end{bmatrix} \quad (2)$$

The ray-transfer matrix $\begin{bmatrix} A & B \\ C & D \end{bmatrix}$ encapsulates the entire optical path:

$$\begin{bmatrix} A & B \\ C & D \end{bmatrix} = \begin{bmatrix} 1 & 0 \\ 0 & \frac{n_2}{n_1} \end{bmatrix} \begin{bmatrix} 1 & d_3 \\ 0 & 1 \end{bmatrix} \begin{bmatrix} 1 & 0 \\ \frac{2}{R_2} & 1 \end{bmatrix} \begin{bmatrix} 1 & d_2 \\ 0 & 1 \end{bmatrix} \begin{bmatrix} 1 & 0 \\ \frac{2}{R_1} & 1 \end{bmatrix} \begin{bmatrix} 1 & d_1 \\ 0 & 1 \end{bmatrix} \begin{bmatrix} 1 & 0 \\ 0 & \frac{n_1}{n_2} \end{bmatrix} \quad (3)$$

Here, $n_1$ and $n_2$ denote the refractive indices of air and polymer, respectively; $d_1$, $d_2$, $d_3$ represent the distances along the optical path, and $R_1$, $R_2$ are the curvature radii at the reflection points, as detailed in Fig 2 (a). Achieving low-loss coupling requires maximizing the overlap integral between the output mode and the fibre mode, demanding an aberration-free condition, defined mathematically by:

$$\begin{cases} x' = Ax + B\theta \\ \theta' = Cx + D\theta \end{cases} \quad (4)$$

To ensure an aberration-free mode, the optical path must satisfy:

$$R_1 + R_2 = 2d_2 \quad (5)$$

This condition is essential to maintain optimal focusing and minimal beam divergence after reflections. By combining the aberration-free and TIR conditions, the design space for reflector geometries is significantly narrowed, substantially reducing computational demands.

This comprehensive theoretical approach provides not only an effective means of optimizing coupler design but also critical insights into the relationship between geometric parameters and optical performance. Such a model substantially enhances design efficiency and reliability, ultimately facilitating the successful implementation of broadband and highly adaptable photonic couplers

*2.2. Numerical simulation*

Lumerical FDTD (finite-difference time-domain) simulations were subsequently conducted to visualize and analyze the light propagation within the reflector system, utilizing the waveguide mode as the input source. These simulations provided critical insights necessary for refining and optimizing the coupler design. To maintain consistency with the theoretical framework previously described, the simulation configuration strictly followed key assumptions, particularly the quasi-parallel beam condition with minimal divergence.

To approximate this idealized input condition in practice, a spherical collimation lens was integrated at the reflector's input facet. As illustrated in Fig. 3 (b), this collimator adopted a truncated spherical geometry with a diameter of 8 μm—slightly larger than the mode field diameter (MFD) of the input optical fiber—to ensure comprehensive capture and effective collimation of the emitted optical mode. To quantitatively evaluate the collimation effectiveness, we assessed the mode overlap between the collimated output and the original fiber mode. Simulations indicated optimal collimation performance and coupling efficiency when the spherical surface radius was set to 15 μm. Furthermore, broadband simulations across the 800 nm to 1670 nm wavelength range demonstrated consistently high coupling efficiency, validating the lens's ability to maintain excellent beam quality and alignment over a super-wide spectral window.

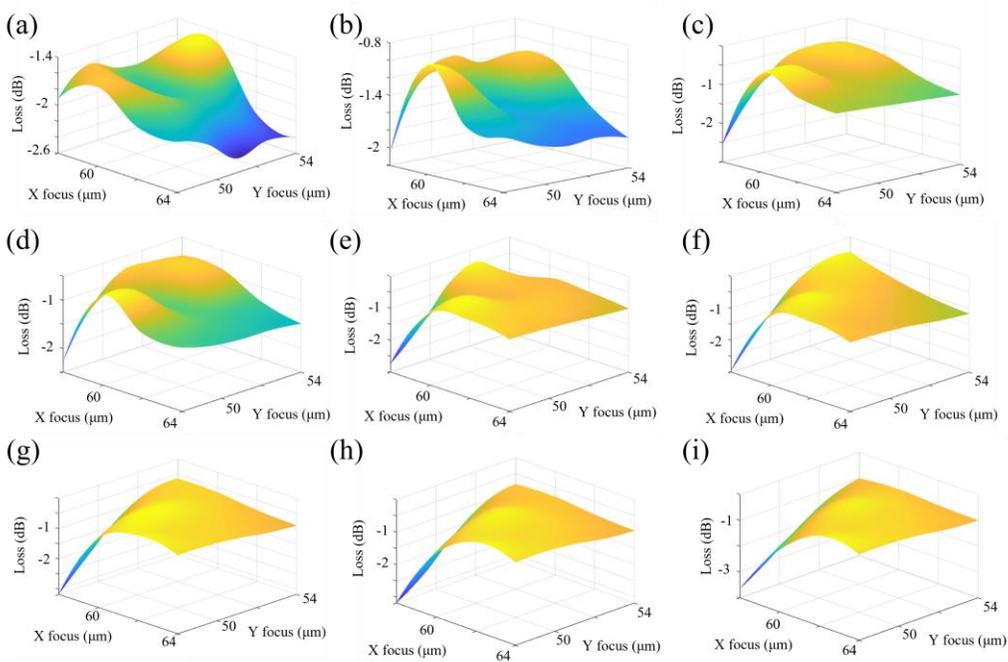

Fig. 3 Simulation results demonstrating fibre-to-fibre coupling losses across different wavelengths, from 800 nm to 1600 nm, shown sequentially in (a) through (i), respectively.

Upon validating the optimal collimator geometry, we initially established a baseline ellipsoidal coupler configuration using identical 980HP fibers as input and output, which serves as a baseline reference for the subsequent optimization of ellipsoidal reflector geometry as illustrated in Figure. 3. Here, the optimization primarily focused on adjusting the foci positions along orthogonal X and Y axes, thereby simplifying simulation complexity and enhanced clarity. The simulated optimal configuration achieves a minimum insertion loss of ~0.5 dB, with the x-axis and y-axis foci are set at 60 μm and 48 μm, respectively. Further broadband simulations reinforced a consistent low-loss operation across 800 nm to 1600 nm. Such stability primarily is attributed to effective aberration management and minimal optical leakage, combining beneficial attributes of guided-wave optics with the flexibility inherent to free-space optical components.

Following the baseline coupler geometry, we then modified the geometry of output reflector to match different mode sizes by fine-tuning the x-intercept and y-focal parameters. For example, we demonstrated our coupler's adaptation from a 980HP fiber mode to an approximately 4 μm mode size. Optimized results yielded an insertion loss of ~0.9 dB over 800 nm to 1600 nm, highlighting the design's robustness and versatility.

## 3. Device fabrication and characterization

To facilitate a fully passive alignment process, an integrated plug-in funnel structure was fabricated enclosing the optical coupler. The funnel allows the stripped optical fiber (with a standard diameter of 125 μm) to be directly and securely inserted into the reflector. The funnel dimensions were precisely designed at 400 μm × 300 μm × 300 μm, featuring a truncated geometry composed of a conical entrance transitioning smoothly into a cylindrical segment.

The funnel's entrance diameter, representing the cone's base, was deliberately set to approximately twice the diameter of the stripped optical fiber (250 μm) to ensure effortless fiber insertion. As the fiber progresses deeper into the funnel, the inner diameter gradually narrows, eventually matching precisely the diameter of the stripped fiber. Accounting for potential resin shrinkage during fabrication, the final internal diameter of the cylindrical segment was accurately set to 130 μm. Additionally, as depicted in Figure. 4 (a), small holes were strategically positioned on the funnel's top surface to facilitate effective resin drainage, enhancing the structure's overall quality and ease of processing.

The fabrication was performed using a commercial direct laser writing system (Nanoscribe Professional GT2), utilizing a 25× objective lens and IP-n162 resin supplied by Nanoscribe. This specific resin was selected for its suitable refractive index (approximately 1.6 at 1310 nm wavelength) and optimal balance between high-resolution capability and efficient fabrication speed. The resin underwent two-photon polymerization (TPP) initiated by exposure to an infrared femtosecond laser, achieving sub-micron precision essential for fine optical structures. To optimize the fabrication process, different laser writing parameters were employed according to the functional requirements of the optical and mechanical components. The optical elements demanded finer slicing and hatching distances coupled with lower laser power to achieve maximum resolution and precision, whereas the mechanical components, such as the funnel, utilized larger slicing and hatching distances with higher laser power to significantly reduce overall fabrication time. Post-exposure processing involved immersing the polymerized structures in propylene glycol monomethyl ether acetate (PGMEA) for 15 minutes to effectively remove residual unexposed resin, followed by rinsing in isopropanol (IPA) for 5 minutes to eliminate any remaining PGMEA. Finally, the fabricated device was subjected to UV curing, significantly enhancing mechanical stability and ensuring structural robustness suitable for practical application.

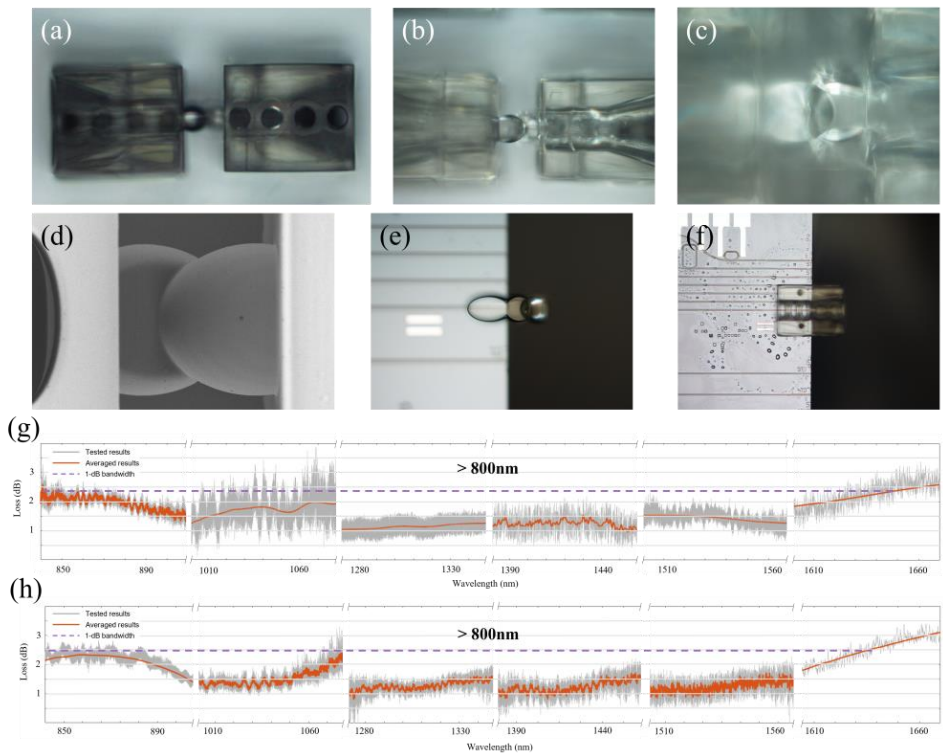

Fig. 4 Images and broadband coupling test results of the fabricated photonic coupler devices. (a)–(c) Optical microscope images illustrating fibre-to-fibre coupling demonstrations; (d) SEM micrograph highlighting detailed structural features of the optical coupler; (e) On-chip integrated coupler structure; (f) On-chip fabricated plug-in funnel containing the coupler; (g) Broadband loss measurement results (840 nm–1670 nm) for the fibre-to-fibre coupling configuration; (h) Broadband loss measurement results (840 nm–1670 nm) for the fibre-to-chip coupling configuration.

*3.1. Fibre-to-fibre demonstration*

As a proof-of-concept demonstration, the coupler was initially evaluated via fibre-to-fibre coupling scheme by using commercially available 980HP and UHNA1 optical fibres from Thorlabs, with an inherent 4:5 mode-size-conversion ratio. Both fibres were stripped to a uniform diameter of 125 μm, matching the dimensions of the co-printed plug-in funnels. The entire freestanding structure, comprising both the plug-in funnel and optical coupler, was fabricated directly onto an ITO-coated substrate, as depicted in Figure 4 (a)-(c). For broadband characterization, superluminescent diodes (SLDs) providing spectral coverage from 800 nm to 1670 nm were employed as the input source. Light emitted from the SLD passed sequentially through the 980HP fibre, the fabricated coupler, and the UHNA1 fibre, before finally being captured and analyzed using an optical spectrum analyzer (OSA).

To optimize fabrication efficiency without compromising structural accuracy, printing parameters—such as slicing and hatching distances—were varied for different device segments. Specifically, optical elements employed fine slicing (100 nm) and hatching (200 nm) distances to maximize precision and minimize optical losses. Conversely, the mechanical plug-in funnel was fabricated using larger slicing (400 nm) and hatching (800 nm) distances, significantly reducing fabrication time. Overall, the fabrication process was completed within approximately 45 minutes, with potential further reductions achievable via scaffold and shell printing techniques. Experimental results demonstrated consistently low coupling losses across the entire measured spectrum from 830 nm to 1630 nm. A minimal loss of 1.2 dB was recorded and the highest loss was about 2.2 dB at approximately 1620 nm wavelength region, indicating a 1dB bandwidth over 800nm range. The noticeable power fluctuations are believed to be caused by residual resins which can be effectively mitigated with an additional annealing step as discussed in Section 3.3 below.

*3.2. Fibre-to-chip coupling test*

Following the fibre-to-fibre coupling case, a fibre-to-chip coupling test was conducted to validate the coupler's performance with PICs. The chip employed in this experiment was fabricated via the CORNERSTONE Photonics multi-project wafer (MPW) run, in a SiN platform with waveguides that have a mode field diameter of approximately 2 μm × 3 μm. By slightly adjusting the geometry of the second ellipsoidal reflector, effective mode matching to this waveguide dimension was straightforwardly achieved. The structural design of the on-chip printed funnel and optical coupler differed slightly from the fibre-to-fibre version. Specifically, to enhance mechanical stability and improve adhesion to the chip surface, part of the coupler structure was printed directly onto the photonic chip, as depicted in Fig. 4 (d)-(f). Before fabrication, the chip was securely attached to a fused silica substrate. The experimental setup utilized the same set of broadband SLD sources as in the preceding tests, ensuring spectral coverage from 800 nm to 1670 nm. The input and output interfaces were both coupled through 980HP fibres. The total coupling loss was accurately determined by measuring the optical power difference between the input and output fibres, subtracting intrinsic on-chip waveguide losses. Experimental results indicated a similar coupling loss compared to the freestanding fibre-to-fibre configuration, with the minimal insertion loss measured at approximately 1 dB and a 1 dB bandwidth spanning over 800 nm wavelength range. This result confirms the versatility and applicability of the designed coupler for integrated photonic systems.

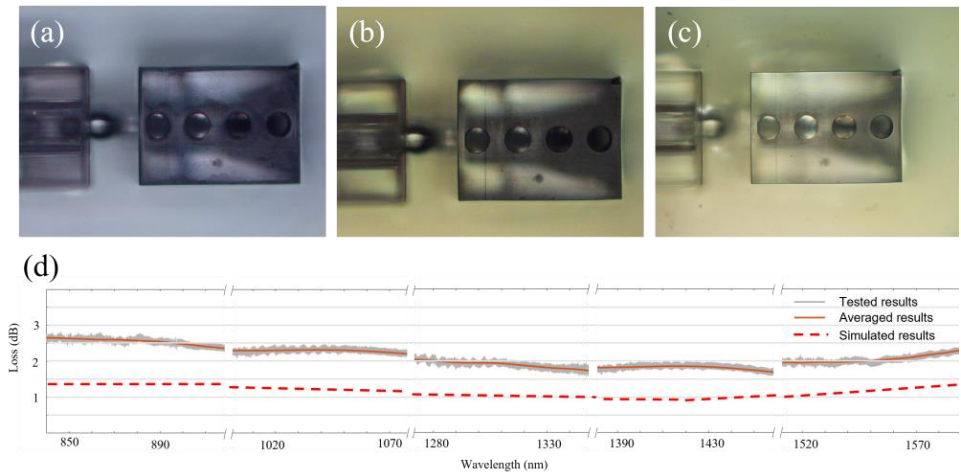

Fig. 5 Images of fibre-to-fibre coupler with funnel under thermal cycles, from (a) to (c): device undergoes 0, 1, 2 thermal cycles; (d) broadband loss test after 2 thermal cycles.

### 3.3. Device thermal robustness test

To systematically evaluate the robustness of our printed device, we performed experiments of thermal-cycling tests using the fibre-to-fibre freestanding assembly. The coupler underwent two programmed cycles in an annealing furnace, each gradual heating up to 150°C over approximately 40 minutes, followed by a controlled cooling phase back to room temperature over approximately one hour. Broadband transmission was measured and recorded after each cycle. As illustrated in Figure 5, the structural integrity of the coupler remained stable, and coupling losses exhibited negligible variations across thermal cycles. The coupling loss gets marginally increasing and converges to a value in between sub-2dB and 2.6dB through the entire measured 740 nm wavelength range. In the meantime, measured power fluctuation decreases, likely caused by the dry-out of residual resin. These findings not only demonstrate the thermal stability of the coupler, but also shed light on the path forwards to lower loss and higher reliability, through an extra thermal annealing step.

### 3.4. Towards complete wavelength-agnostic coupling

The designed coupler has delivered an exceptional broadband performance but an increased loss persists at the edges. This is attributed to the residual chromatic aberrations present in the current ellipsoidal geometry, but is also an outcome of deliberate design trade-offs balancing performance and complexity. Looking ahead, for a complete wavelength-agnostic coupler, global optimization strategies, including advanced machine-learning-assisted inverse design methods can be used, to systematically explore larger, non-intuitive parameter spaces. Such methods could achieve superior minimization of spectral aberrations by identifying optimal geometric configurations that conventional heuristic or analytical approaches overlook. Specifically, the quantization of the continuous reflector contours would provide enhanced precision and control over aberration management, as indicated by Fig. 6, enabling finer tuning of the coupler's optical performance. In addition, fabrication-induced imperfections, notably nanoscale surface roughness and discretization errors associated with curved reflector geometries, are significant contributors to residual insertion losses observed at certain wavelengths. Advanced fabrication methodologies, including enhanced printing resolution and sophisticated post-processing treatments, can be applied to mitigate these issues.

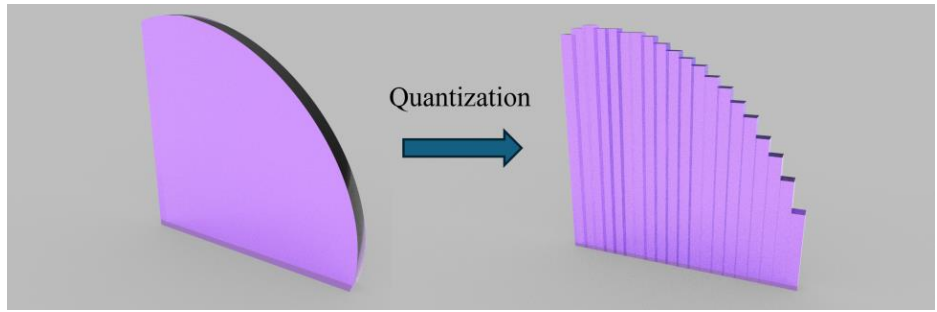

Fig. 6 Possible methods to further broaden the coupler's operating bandwidth. By quantizing the ellipsoidal reflector geometry, the output beam profile can be more precisely engineered, paving the way towards truly wavelength-agnostic optical coupling.

## 4. Conclusion and Outlook

In this work, we present a wavelength-agnostic photonic coupler via 3D nano-printing. Featuring an innovative dual-ellipsoidal geometry that facilitates quasi-free-space optical propagation, our coupler achieves an insertion loss as low as 1.3 dB across a record-wide 1 dB spectral bandwidth exceeding 800 nm. This design demonstrates not only high coupling efficiency, but also excellent mechanical robustness and stability under diverse operational conditions. The fabrication process adopts a single-step 3D printing strategy that ensures strong adhesion to a wide range of photonic platforms—from mature integrated systems such as silicon-on-insulator (SOI) and silicon nitride (SiN), to emerging polymer-based flexible platforms—enabling compatibility with standard post-processing and packaging workflows.

While the current implementation primarily targets single-polarization transverse electric (TE) modes, future work will focus on enabling polarization diversity, including efficient coupling of transverse magnetic (TM) modes and even support for multimode operations. Moreover, as the optical reflector occupies only a small portion of the total structure, multiple couplers can be spatially converged and overlapped to enable high-density, multi-port photonic interconnections. Overall, the demonstrated flexibility, scalability, and material adaptability highlight the strong potential of our design for deployment across diverse broadband photonic technologies, representing a key advance toward next-generation high-performance photonic interfacing.

**Funding.** This work was funded by the UK Research and Innovation, Engineering and Physical Sciences Research Council (UKRI-EPSRC), project QUDOS (EP/T028475/1); and the European Union's Horizon Europe Research and Innovation Program, project PUNCH (101070560) and project INSPIRE (101017088).

**Disclosures.** The authors declare no conflicts of interest.